\documentclass[twocolumn]{aa}  
\usepackage{amsmath}
\usepackage{graphicx}
\usepackage[usenames]{color}
\usepackage{natbib} 
\bibpunct{(}{)}{;}{a}{}{,} % to follow the A&A style 
\usepackage{multirow}
\usepackage[breaklinks,colorlinks=true,citecolor=black,linkcolor=black,urlcolor=blue,bookmarks=true]{hyperref}
\usepackage{breakurl}  % to break long URL on multiple lines

\newcommand{\second}{$^{\prime\prime}$}
\newcommand{\kms}{km\,s$^{-1}$}
\newcommand{\ulyss}{\href{http://ulyss.univ-lyon1.fr}{\textcolor{black}{\emph{ULySS}}}}
\newcommand{\teff}{$T_{\rm eff}$}
\newcommand{\logg}{log($g$)}
\newcommand{\csq}{$\chi^2$}
\newcommand{\ie}{i.\,e.,}
\newcommand{\eg}{e.\,g.,}

\usepackage{txfonts}
\begin{document}
   \title{\emph{ULySS}: A Full Spectrum Fitting Package}

%   \subtitle{...}

   \author{
          M. Koleva\inst{1,2},
          Ph. Prugniel\inst{1},
          A. Bouchard\inst{1,3}
          \and
	  Y. Wu\inst{1,4}
          }

   \offprints{M. Koleva, \email{ulyssobs.univ-lyon1.fr}}
   \institute{%
Universit\'e de Lyon, Lyon, F-69000, France ; Universit\'e Lyon~1,
Villeurbanne, F-69622, France; Centre de Recherche Astronomique de
Lyon,
Observatoire de Lyon, St. Genis Laval, F-69561; CNRS, UMR 5574; 
Ecole Normale Sup\'erieure de Lyon, Lyon, France
 \and
Department of Astronomy, St. Kliment Ohridski University of Sofia, 5 James
Bourchier Blvd., BG-1164 Sofia, Bulgaria
 \and
Department of Astronomy, University of Cape Town, Private Bag X3, Rondebosch 7701, South Africa
 \and
National Astronomical Observatories, Chinese Academy of Sciences, 20A Datun Road, Chaoyang District, 100012, Beijing, China
}

   \date{\today}

  \abstract
  % context heading (optional)
 {} %leave it empty if necessary  
   {
    We provide an easy-to-use full-spectrum fitting package and
    explore its applications to (i) the determination of the 
    stellar atmospheric parameters and (ii) the study of the history 
    of stellar populations.
   }
  % methods heading (mandatory)
   {We developed \emph{ULySS}, a package to fit spectroscopic observations against a linear 
    combination of non-linear model components convolved with a parametric 
    line-of-sight velocity distribution. The minimization can be either local or global, and determines
    all the parameters in a single fit.
    We use \csq{} maps, convergence maps and Monte-Carlo 
    simulations to study the degeneracies, local minima and to
    estimate the errors.
}
  % results heading (mandatory)
   {
   We show the importance of determining the shape of the continuum
   simultaneously to the other parameters by including a multiplicative
   polynomial in the model (without prior pseudo-continuum determination, or
   rectification of the spectrum). We also stress the benefice
   of using an accurate line-spread function, depending on the wavelength,
   so that the line-shape of the models properly match the observation.  
   For simple models, i. e., to measure the atmospheric parameters or
   the age/metallicity of a single-age stellar population, there is often
   a unique minimum, or when local minima exist they can unambiguously
   be recognized. For more complex models, Monte-Carlo simulations
   are required to assess the validity of the solution.
}
%  % conclusions heading (optional), leave it empty if necessary 
   {The \emph{ULySS} package is public, simple to use and flexible. The full 
   spectrum fitting makes optimal usage of the signal.}


\maketitle

\section{Introduction}
Spectroscopic data from astronomical sources contain most of the information
that we can get from the remote universe, such as physical conditions, 
composition or motions.
Often, the information extraction relies on the identification 
and analysis of spectral signatures due to more or less well understood physical 
processes. 
This is usually an interactive task and requires a specialised 
expertise, although this is becoming less true with time \citep[\eg{}][]{sousa07}.  

The recent data avalanche
prompted a search for automated, objective (and generally more efficient) 
methods.One 
possibility is to directly compare an observed spectrum with a set of models 
or an empirical library of object with known characteristics on a pixel to pixel 
basis. This can be done for the analysis of stellar atmospheres as in 
\citet{bailer-jones97}, \citet[TGMET]{katz98}, \citet{prugniel01}, \citet{snider01}, 
\citet{willemsen05}, \citet{shkedy07} and \citet[MATISSE]{MATISSE}. 
It has also been used for the determination of the history of stellar populations as in 
\citet[MOPED]{MOPED}, \citet{panter03}, \citet[Starlight]{cidfernandes05}, 
\citet{moultaka05}, \citet[STECKMAP]{ocvirk06a, ocvirk06b} and 
\citet[VESPA]{VESTA}.

The main advantage of these methods is that, rather than picking specific 
features, they make optimal usage of the entire measured signal. This comes with 
a price, however, as we forfeit our ability to draw direct relations between 
the strength of a specific feature and a physical characteristic. This lack of 
simplicity should ultimately be regarded as inherent to the nature of the 
spectra, where the information is redundant and distributed (possibly uniformly) 
over a large wavelength range.

This paper presents \ulyss{} (Universit\'e de Lyon Spectroscopic analysis 
Software)\footnote{\ulyss{} is available at: \url{http://ulyss.univ-lyon1.fr/}}, 
a new software package enabling full spectral fitting for two 
astrophysical contexts: The determination of (i) stellar atmospheric parameters and 
(ii) star formation and metal enrichment history of galaxies. Many similarities 
between these two cases allowed us to build a single package capable of handling both. 

In \ulyss{}, an observed spectrum is fitted against a model expressed as
a linear combination of non-linear components, optionally convolved 
with a line-of-sight velocity distribution (LOSVD) and multiplied with
a polynomial function. A component is a non-linear function
of some physical parameters (\eg{} \teff, \logg{} and [Fe/H]).
The multiplicative polynomial is meant to absorb errors in 
the flux calibration, Galactic extinction or any other cause affecting 
the shape of the spectrum. It replaces the prior {\it rectification} or 
normalization to the pseudo-continuum that other methods requires.
This model is compared to the data through a non-linear least-square
minimization.

\ulyss{} has been used to study stellar populations 
\citep{koleva08a, michielsen07, koleva08b, koleva09, bouchard09} and
determine atmospheric parameters of stars
\citep{prugniel09}.

The goal of this paper is to describe the package
and to explore the potential and caveat of  
direct comparison between an observation and a composite model. 
A previous implementation of the same idea, shortly described earlier
and named NBURSTS \citep{koleva07,koleva08a}, was a variant of the pPXF program
 \citep[hereafter CE04]{cappellari04} applied to stellar populations.
Another variant of pPXF was developed by \citet{sarzi06} to fit in the
same time the stellar and the gas kinematics.
The main reason for the name change is the widening of the scope of the 
package to the measurement of atmospheric parameters
and potentially other applications.
In addition the algorithm has been modified in some of its mathematical 
details to improve its precision, robustness and performance.
We also made the package user-friendly, wrote documentation and tutorials
to allow the public distribution of the package.

The paper is organized as follow: In Sect. 2 we 
describe the method and the package. In Sect. 3 we illustrate the 
approach with the case of the determination of the atmospheric 
parameters of a star. In Sect. 4, we give examples of analysis 
of a stellar population. The Sect. 5 draws conclusions.

\section{Description of the method and package}

The method consists in minimizing the \csq{} between an 
observed spectrum and a model. The model is generated at the same 
resolution and with the same sampling as the observation and the fit 
is performed in the pixel space. This optimizes the usage of the 
information, as all the signal is used, and allows easy consideration 
of the errors on each spectral bin.

The method proceeds in a single fit to determine all the free parameters
in order to handle properly the degeneracies between them
(\eg{} the temperature-metallicity for a 
stellar spectrum).
Other methods estimate the parameters in different steps. For example,
one may measure the temperature using criteria {\it almost} insensitive
to the metallicity and in turn, using this temperature, derive the 
metallicity. If in fact the criteria to obtain the temperature is
somehow metallicity-dependent, the absolute minimum will not be reached
unless the fit is iterated. Using a single minimization do not require
this additional complexity.

\citet[][hereafter STECKMAP]{ocvirk06a,ocvirk06b} presented another 
implementation of this approach, using a non-parametric regularized fit. 
It performs very well to reconstruct the evolution history of a stellar 
population. But one of its limitation is the 
difficulty to grant a degree of confidence on the solution
(and it is tailored to stellar populations only). This 
is the main reason which lead us to work on a `simpler' parametric 
minimization allowing to better understand the degeneracies and 
the structure of the parameter space by constructing \csq{} maps.

The parametric local minimization
presented here has been introduced and compared with STECKMAP in
\citet{koleva08a} in the case of single stellar populations.

\subsection{Parametric Model}

The observed spectrum, $F_{obs}(\lambda)$, is approximated by a linear 
combination of $k$ non linear components, CMP, each with weight $W$. 
This composite model is possibly convolved with a LOSVD and multiplied 
by an $n^{th}$ order polynomial, $P_n(\lambda)$, and summed to another 
polynomial, $Q_m(\lambda)$:

\begin{align}
F_{obs}(\lambda) = P_{n}(\lambda) \times \bigg(
    &{\rm LOSVD}(v_{sys},\sigma, h3, h4) \nonumber \\
    &\otimes \sum_{i=0}^{i=k} W_i \,\, {\rm CMP}_i\,(a_1, a_2, ...,\lambda) 
\bigg) + Q_{m}(\lambda) \label{eq:main}
\end{align}

\noindent The LOSVD is a function of systemic velocity, $v_{sys}$, 
velocity dispersion $\sigma$ and may include Gauss-Hermit expansion 
($h3$ and $h4$, \citealp{marel93}). $\lambda$ is the logarithm of the 
wavelength (the logarithmic scale is required to express the effect of 
the LOSVD as a convolution). The CMP$_i$ must be tailored to
each problem. For instance,
to study a stellar atmosphere, the CMP will be a function of
the effective temperature, \teff, surface gravity, $g$, and
metallicity, [Fe/H]. Or, for a stellar population, it will
be a function of age, [Fe/H], and [Mg/Fe],  returning the spectrum
of a single stellar population (SSP).
In general, a CMP is a function of an arbitrary number of physical
parameters and of $\lambda$.

The importance of the multiplicative polynomial has been discussed
in \citet{koleva08a} as it absorbs the effects of an unprecise
flux calibration (a common issue in small-aperture spectroscopy)
and of the Galactic extinction (which could also be explicitly
included in the model as in STECKMAP).
The effect of  $P_n(\lambda)$ is studied in Sect.~\ref{sect:mulpol}
in the particular case of the determination of atmospheric parameters
of a star. A similar study was made in \citet{koleva08a} in the
case of stellar populations.
In all the practical cases 
that we studied, $P_n(\lambda)$ was  not degenerate with the 
parameters of the CMP, and high orders could be used
(though $n\,\approx\,10$ is often sufficient to obtain an
unbiased estimate of these parameters). The optimal value of $n$, which 
depends mostly on the resolution and wavelength range, can be chosen 
with \ulyss{}.

The additive polynomial is certainly of a more delicate usage, 
and is, in most cases, absent from the model. 
It is indeed degenerate with the intensity, depth or equivalent width 
of the lines and may bias determinations of the CMP parameters 
(\teff, age or [Fe/H], depending on the context).
Such a term may be included to account for data-processing
errors (under- or over-subtraction of a smooth background),
or may have a physical origin.
When determining the stellar kinematics by fitting one or several 
constant  CMPs (\ie{} fixed spectra, like stars or models without 
free parameter) the additive term may be required to absorb the 
template mismatch \citep{koleva08c}. 
Omitting the additive
term would in that case bias the measurement of the velocity dispersion.
For this reason, an additive term is generally present in the 
packages used to determine the internal kinematics of a stellar 
population. For instance, in CE04 it is 
an explicit polynomial, or in Fourier quotient programs \citep{sargent77}
the spectrum is continuum- or mean-subtracted before the fit and the 
amplitude or the features is a free parameter (this is equivalent to an 
additive term).
Another place needing an additive term is the analysis of
a stellar population in presence of the nebular continuum emission of an AGN 
(in that case a power-law is more appropriate than a Legendre polynomial;
see the tutorial distributed with the package). The
biases on the stellar population parameters may be determined
from simulations \citep{Chilingarian05,Prugniel07c}.

In summary, we can only advise careful usage of the additive terms. They are only needed in rare circumstances and including them without valid reasons (or failing to do so when they are required) may severely bias the results of the analysis.
The lowering of the \csq{} that is likely to result from the inclusion
of additional degrees of freedom should not be the only criterium to 
evaluate the validity of the model.
The user should check the effects of these terms on the CMP parameters using simulations.

\subsection{Algorithm}

As written in Eq.~\ref{eq:main}, the problem is a fit to a multilinear 
combination of non-linear functions. The parameters of each CMP$_{i}$ 
are in general non-linear and are evaluated together as those of the 
LOSVD with a Levenberg-Marquardt \citep{marquardt63, more80} routine 
(hereafter LM).
To measure the $h3$ and $h4$ coefficients of the LOSVD the package
implements the penalization proposed by CE04 to bias 
them to 0. This option, not illustrated here, is 
useful to analyse low S/N data.
 
The (linear) coefficients of the  $P_n(\lambda)$ and  $Q_m(\lambda)$ 
polynomials are determined by ordinary least-square at each
evaluation of the function minimized by the LM routine.
The weights of each components, $W_i$, are also determined
at each LM iteration using a bounding value least-square
method \citep{lawson95} in order to take into account constraints on the
contribution of each component, like imposing positivity.

An alternative would have been to use the variable projection method
\citep{golub73}
for solving this separable nonlinear least-squares problem
(i. e. where the model is a linear combination of non-linear functions).
We did not use this solution because the bounding of the linear parameters
is not possible in the public implementation of this method
\cite[the VARPRO program]{bolstad77}, in addition there was no version of 
this algorithm in the IDL/GDL language used for this project. Therefore
we preferred to adjust the linear parameters in a separate layer
at each LM iteration, as in \citet{cappellari04}. We stress that separating
the linear variables is important, not only for performance reasons
but also for stability.

The bounding on the $W$ is an option that can be tuned on the
particular case of a $CMP$. For example, when dealing with the 
decomposition of a stellar population as a series of bursts,
each component must have a positive or null weight.
The LM implementation by 
\citet[in MINPACK\footnote{\url{http://www.netlib.org}}]{more80}
gives the freedom to also define limits for the values of
any of the parameters of the $CMP$.

For $P_n(\lambda)$ and $Q_m(\lambda)$, we use Legendre polynomial 
developments of respectively orders $n$ and $m$, 
for their orthogonality properties. 
However, the developments intervening in our problem are
not strictly orthogonal because 
(1) the errors are not uniform along the $\lambda$ axis and 
(2) there may be gaps in the signal corresponding to masked pixels (missing
or damaged values).
For these reasons, the $P_n(\lambda)$ and  $Q_m(\lambda)$ are determined by
an ordinary least square fit. Defining ad'hoc orthogonal polynomials
would have been equivalent both from mathematical and performance
point of views.

Note that as the LM minimization makes its path to the solution using
the local gradients of the CMPs, there is no need or advantage to apply
a linear transformation to these functions. 
For example, applying a rotation in the age -- metallicity plan in order
to minimize on two orthogonal parameters does not ease the convergence
or avoid local minimas: The path to the solution is not affected.
But a proper choice of the parameters may lead to a better convergence.
For instance, to fit a stellar spectrum,
minimizing on log(\teff) or on $\theta=5040/$\teff{} is preferrable
than using directly \teff.

\subsection{Line Spread Function}

To compare a model to an observation, both must have the same spectral
resolution, or we must first transform either the model or
the observation in order to match their respective resolution.

The spectral resolution is characterized by the line spread function, LSF,
which is the analogous in the spectral direction of the PSF
(point spread function) for images. This is the {\it impulse response}
describing the wavelength distribution of the flux of an un-resolved
spectral line. The LSF results from the convolution between the
intrinsic resolution of the spectrograph and the slit function.
In first approximation, it is represented by $R = l/\Delta(l)$,
where  $l$ is the wavelength (linear scale) and 
$\Delta(l)$ is the FWHM of the LSF.

In practice, the LSF may not be defined by a single number. It is
not necessarily a Gaussian, and may change with wavelength
and position in the field for integral-field or long-slit
spectroscopy. Usually, the model has a higher resolution than the
observation. 
Otherwise, the
analysis will not make optimal usage of the available information 
(\ie{} the high resolution details in the observed spectrum will not
be exploited). 

To make the model comparable to the observations, we proceed in two
steps. First we have to determine the LSF and then {\it inject}
it in the model.

\subsubsection{Determination of the line-spread function}

\begin{figure*}
\includegraphics{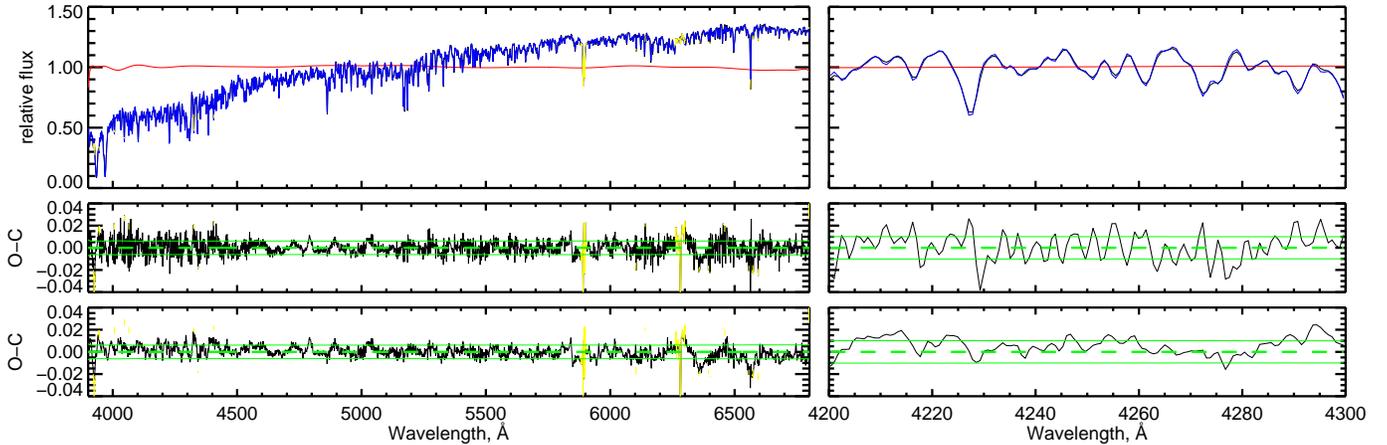}
\caption {Effect of using a precise LSF, illustrated with a fit of a 
Vazdekis/Miles spectrum with a Pegase.HR/Elodie.3.1 SSP component.
The top panel shows the spectrum (in black) and the best fit in blue 
(both are almost superimposed and the black line can be seen only when 
zooming on the figure), the red line is the multiplicative polynomial. 
The yellow regions where rejected from the fit (rejection of flagged 
telluric lines and automatic rejection of outliers). The middle and bottom 
panel are respectively the residuals from the best fit when (i) assuming
a constant Gaussian LSF (in $\lambda$) or (ii) a matched LSF. The continuous 
green lines mark the 1-$\sigma$ deviation, and the dashed line is the zero-axis.
The right side of the graphic expands a small wavelength region, around Ca4227.
}
\label{fig:lsf1}
\end{figure*}

The function that we are seeking is the {\it relative} LSF between
the model (which have a finite resolution) and the observation. 
It should normally be determined using calibration observations.

Three types of calibrations can be considered:
\begin{enumerate}
\item
Arc lamp spectra. They are routinely produced during the observations
and are used to adjust the dispersion relation and to achieve the
wavelength calibration. The slit of the spectrograph 
is uniformly illuminated with a discharge lamp (like for example He-Ar) 
producing narrow emission lines. The position of chosen unblended
lines are used to fit the dispersion, and the width and shape can be
used to determine the LSF.
\item
Standard star. Normally any star, except some hot stars with
featureless spectra used for the spectrophotometric calibration, 
can be used to determine the LSF. The observed spectrum may
be compared with a high-resolution spectrum of the same star,
or with a model of this star, to determine the broadening
due the the spectrograph. \ulyss{} can naturally be used
to measure this broadening.
(Beware that sometimes spectrophotometric standards
are observed with a widened slit, and are not usable for
LSF calibration).
\item
Twilight spectrum. Spectra of the twilight sky are often used
to determine the variation of the sensitivity over the field 
of the spectrograph. These spectra result from the uniform
illumination of the slit by a Solar spectrum and can therefore
by used as any stellar spectrum to measure the broadening.
\end{enumerate}

The first solution, with arc spectrum, may appear simpler, as
it contains bright and well separated emission lines that can be
individually fitted with a Gaussian or Gauss-Hermite development.
However, there are some caveat with this approach: (i) Few lines are completely 
unblended and profiles are sensitive
to faint unresolved neighboring lines; (ii) The lines
are often bright, and use the detector in a regime strongly 
different than the observations of the astronomical sources,
therefore their profile may be affected by some small 
non-linearities; 
(iii) The spectrographs are often used close to the
undersampling limit (the width of the arc lines are about 2 pixels)
and fitting a profiles in these conditions is perilous; 
(iv) The illumination of the slit is not
exactly the same as for the astronomical sources (different optical paths), 
and (v) this method determines the {\it absolute} LSF that needs to
be deconvolved from the models's LSF before using it.
We recommend to use this solution only as a sanity control.

The two other solutions use stellar spectra. As the physical
models presented in this article are based on empirical 
libraries of stellar spectra, a proper choice of the 
calibration stars can make the task of determining the LSF 
straightforward (the solar spectrum is included in
most libraries). If the exact star is not available
at the model's resolution, a similar star, or an
interpolated spectrum may be used.
Using a stellar spectrum bypasses some of the difficulties
met with arc spectra and can gives directly the relative LSF.

There is still an important phenomenon to consider when
describing the LSF. Often, the intrinsic resolution of the
spectrograph is significantly higher than the actual resolution
which is limited by the slit width. As a consequence, the distribution 
of light within the slit has an effect on the 
spectrum. In particular, the apparent broadening of a star
observed under excellent seeing condition (seeing smaller than
the slit) will be smaller that the broadening observed for an
extended object (or a star with poor seeing condition). The
{\it effective} resolution results from the product of the
light profile of the object by the slit function, convolved
by the intrinsic resolution of the spectrograph.

This effective resolution depends on the observing conditions
and on the light profile of the source. It may vary between
consecutive exposures. This problem may be difficult to 
correct and by limiting the knowledge of the instrumental
broadening it hampers the measurement of the physical velocity 
dispersion. In the types of analysis discussed in this paper,
this effect preserves the determination of the other parameters
(metallicity, age or \teff).

Hints for this effect of resolution of an object within the slit 
may be obtained when comparing the LSF derived with various 
standard stars and those obtained with twilight spectra.

A practical mean to measure the LSF is to determine the 
broadening function (cz, $\sigma$ and possibly $h3$ and $h4$)
in a succession of small wavelength intervals. For
spectra with R = 1000 to 3000, we typically use segments
of 200 \AA, separated with 100 \AA{} steps (they overlap by half
their length).

\subsubsection{Injection of the LSF in the model}

Because the LSF varies with wavelength, it cannot be injected in the model
as a mere convolution. The method we use consists in convolving the models by
the series of LSFs determined at some wavelength and then make a
linear interpolation in wavelength between the convolved models.

To illustrate the importance of matching the LSF in the spectrum
fitting process, we show in Fig.~\ref{fig:lsf1} the analysis of a spectrum
with a Pegase.HR population model based on the Elodie.3.1 library.
The analysed spectrum is
a stellar population model from the library of SSPs computed by 
Vazdekis\footnote{\url{http://www.iac.es/galeria/vazdekis/vazdekis_models_ssp_seds.html}} 
with the Miles library \citep{sanchezblazquez06}.
The first fit simply assumes a purely Gaussian and constant
LSF. The second uses an optimal LSF.
The residuals of the first fit are minimal in the center, but
become larger at the edges, a zoom in a small region
shows that this is due to a misfit (not to noise). Using
the proper LSF corrects this defect and the residuals become smoother.
In this particular example, despite the considerable effect on the residuals,
the incidence on the stellar population parameters is marginal.
The precision of the LSF mostly affects the parameters of the LOSVD.
By suppressing the LSF mismatch,
we can search for other signatures in the residuals which
could have been masked otherwise.

The LSF injection also corrects possible inaccuracies in the wavelength 
calibration. An example of this is shown in \citet{koleva08a},
where a wavelength calibration systematic distortion affecting
the \citet{bruzual03} models is corrected.

\begin{figure}
\includegraphics{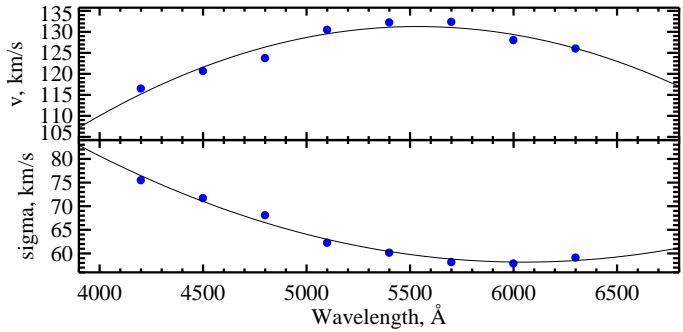}
\caption {Line spead function of the SDSS.}
\label{fig:lsf2}
\end{figure}

\subsubsection{Example: The SDSS LSF}

As an example of LSF analysis, we use the velocity dispersion template
stars from the SDSS copied from
\url{http://www.sdss.org/dr5/algorithms/veldisp.html}.
These 32 G and K giant stars from M67 were used to determine the 
velocity dispersion of the galaxies as an
average between estimates obtained by Fourier-fitting and 
direct-fitting methods.

In Fig.~\ref{fig:lsf2} we show the LSF relative to the Elodie.3.1
library obtained with \ulyss{} (Elodie.3.1 is restricted to the wavelength
range 3900--6800 \AA).
It was determined using wavelength intervals of 600 \AA{} spaced by 300 \AA{}.
The variation of the instrumental velocity dispersion ($\sigma_{ins}$) 
with wavelength is significant: from 50 to 75 \kms.

The classical methods for measuring the (physical) velocity dispersion, $\sigma$,
\citep{sargent77,tonry79,franx89,bender90,rix92, marel93, cappellari04}
compares the observation to stellar templates observed with
the same setup. As in general, the LSF changes are moderate, 
the red-shift of the galaxy does not affect significantly $\sigma$
for nearby galaxies. But for a distant galaxy, neglecting the
shift of the LSF may give a measurable effect. For a low-mass 
galaxy with $\sigma$ = 50 \kms{}
measured with the SDSS, the bias would be about 0.1 \kms{} at z=0.03
but 2 \kms{} at z=0.4.

\subsection{Description of the \ulyss{} package}

\ulyss{}, available at \url{http://ulyss.univ-lyon1.fr/}, has been 
programmed in IDL/GDL language. This is the language of the widely used proprietary 
IDL (Interactive Data Language)\footnote{\url{http://www.ittvis.com/idl/}} software.
The open source GDL (Gnu Data Language)\footnote{\url{http://gnudatalanguage.
sourceforge.net/}} interpretor can also be used to run \ulyss{}. The choice of 
this programming language is essentially historical: Many of the required routines 
were already available as public libraries and development was started from the 
pPXF package \citep{cappellari04}. We may offer a version written in a modern 
language (most likely Python) in the future.

\ulyss{} contains various programs and subroutines that can be used for:
\begin{itemize}
\item Defining the array of components to fit
\item Performing the fit
\item Visualising the results
\item Making \csq{} maps, convergence maps and Monte-Carlo simulations
\item Reading data from FITS files
\item Testing the package
\end{itemize}

The package contains extensive documentation and tutorials. The emphasis was
put on the flexibility and ease of use.

The package contains routines
to define various types of CMPs, notably to analyse a stellar atmosphere (TGM)
and to analyse a stellar population (SSP), and the construction of other CMPs
by the user was made as easy as possible. To fit a composite model, \ie{} a
linear combination of components, one simply has to concatenate several
CMPs in a single array.

The core of the package is the local minimization described in Sect.~2.
Such a minimization starts from a point (guess) in the parameters space, 
whose choice may be critical. To release this constraint, \ulyss{} makes it easy
to perform a {\it global} minimization by providing vectors instead
of scalars as guesses.

The most valuable aspect of the package is the possibility to explore
and visualize the parameters space. The tools offered for that purpose
are (i) Monte-Carlo simulations (ii) convergence maps, and
(iii) \csq{} maps.

Monte-Carlo simulations are performed to estimate the biases, the errors and the
coupling (degeneracies) between the parameters. A simple option in the
main fitting program allows to perform a series of minimizations with
random noise added to the data. This noise has the same characteristics
as the one estimated in the data. Normally the noise should be
carried throughout the data-reduction process, starting from the
statistical noise of the detector that can usually be securely 
estimated. During the data reduction, the signal is likely to
be resampled, when the spectra are extracted from the initial
2D frame, and when they are calibrated in wavelength (or logarithm 
of wavelength). This operation introduces a correlation between the
pixels (see \citealp{debruyne03}), which is represented in \ulyss{} 
by the ratio between
the actual number of pixels and the number of independent pixels.
Using it, the Monte-Carlo simulations reproduces the correct 
noise spectrum and gives a robust estimates of the errors.

Convergence maps are tools to evaluate the convergence region, \ie{}
the domain of the parameter space from which guesses converge to
the absolute minimum of the \csq{}. These maps can be generated
using the global minimization approach explained above.

\csq{} maps are visualizations of the parameters space.
They are generated by (i) choosing a 2D projection 
(\eg{} age and metallicity) and a node grid in this plan and (ii) 
performing an optimization over all the other axes of the parameters
space for each node of this grid. These maps allow the identification
of degeneracies and local minima.

Typically, when approaching a new problem, like using a new CMP, or 
a new wavelength range or region of the parameters space, these three
tools can be used to understand the reliability of the results
before proceeding to the analysis of a massive dataset. Their
usefulness are presented in the next sections.

\section{Determination of stellar atmospheric parameters}

The effective temperature, \teff, surface 
gravity, \logg, and metallicity, [Fe/H], are fundamental
characteristics that can be derived from spectroscopic
analysis \citep{cayrel01}. Full spectrum fitting, as provided by \ulyss{},
can be used for this purpose: The program will identify the
best matching  \teff, \logg{} and [Fe/H] by comparing to a model.

\ulyss{} does a parametric minimization. So, the core of the problem
is to obtain a parametric model, i.e. a function returning a spectrum
given a set of atmospheric parameters. The reference spectra
are either a grid of theoretical models
(e. g. \citealp{munari05,coelho05})
or a set of observed stars whose parameters are known from
the analysis of individual high resolution spectra
(e. g. \citealp{soubiran98}). \ulyss{} requires an
{\it interpolator} of this grid. In the present paper, we
use the one presented in \citet{prugniel01}
for the ELODIE library. Shortly, it consists in polynomial 
approximations of the library. Three overlapping ranges of \teff{}
are considered (hot, warm and cold) and linearly interpolated to
produce the final function. In each \teff{} range each pixel
of the spectrum is computed as a 21 terms polynomial 
in \teff{}, \logg{} and [Fe/H]. The coefficients of these 
polynomials were fitted over the 2000 spectra of the library.
The choice of the terms, of
the \teff{} limits and of weights were fine-tuned to
minimize the residuals between the observations and the 
interpolated spectra. In this paper, we use the
interpolator  built on the continuum normalized spectra.
An alternative solution to this global polynomial representation
of the library, would have been to use a local approximation
based on a gaussian-kernel smoothing, as in \citet[Appendix B]{vazdekis03}.

\ulyss{} defines a model component (CMP in Eq. \ref{eq:main})
for this model. The TGM component, as we named it, allows
to perform the minimization on the three atmospheric parameters.
With the current version of the ELODIE library 
(\citealp{prugniel07}, version 3.1) the temperature range
is limited to
3\,600\,K$<$\,\teff{}\,$<$30\,000\,K. In the future versions 
\citep[][in preparation]{wu09}, 
a greater number of hot (\teff{}\,$>$\,10\,000\,K) and cold 
(\teff{}\,$<$4\,200\,K) stars
will be included, extending the current validity range of the interpolator.

\subsection{TGM Fit Example}

\begin{figure}
\centering
\includegraphics[scale=1.05]{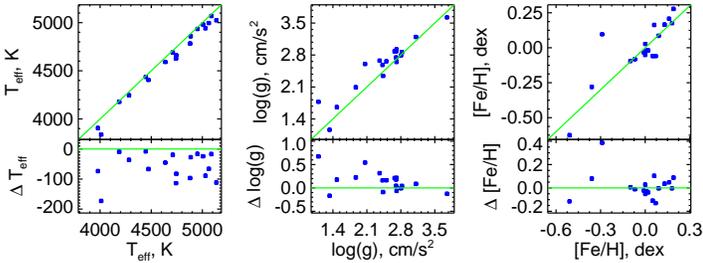}
\caption {Comparison of the atmospheric parameters determined by \ulyss{} with those 
from high resolution spectra \citep{Silva06}.
The abscissas are the measurements from \citet{Silva06}. On the top panels, the
ordinates are from \ulyss{} and the green lines are the diagonals.
On the bottom panels the ordinates are the differences \ulyss{}- literature. 
}
\label{tgm_omc}
\end{figure}

We analysed the 18 stars from the CFLIB \citep[indo-US, ][]{valdes04}
library of spectra in common with the study by \citet{Silva06} of G \& K stars
using R$\approx$50000 spectroscopy and LTE models.
We performed a fit with a TGM component
starting from a grid of guesses in order to be independent of the
prior knowledge of the parameters.
The LSF was determined by using several stars  
in common between CFLIB and the ELODIE library.
The results found with \ulyss{} are consistent with those of \citet{Silva06}, 
Fig.~\ref{tgm_omc}, except for HD~189319 where we find a significant discrepancy 
in metallicity.
The measurements from \citet{Silva06} are: 
\teff=3978 K \logg=1.10 g.cm$^{-2}$, [Fe/H]=-0.29 
and from \ulyss: 
\teff=3904, \logg=1.77, [Fe/H]=0.10. 
It is the most discrepant star for both metallicity
and gravity; it is also the lowest gravity and coolest star of this sample.
Another recent spectroscopic analysis gives:
\teff=4150  \logg=1.70  [Fe/H]=-0.41 \citep{Hekker07}; and interferometric
measurements tend to indicate a lower \teff: 3650 -- 3800 K, and hence probably
lower gravity \logg = 0.9 \citep{Neilson08,Wittkowski06}.
This discrepancy is therefore not raising suspicion concerning \ulyss, but
the ELODIE interpolator surely merits to be improved in this region of the HR diagram.

For this sample
we found 
$\Delta(T_{eff})/T_{eff} = -0.013 \pm 0.010$ (\ie{} $60 \pm 45 $K), $\Delta(log(g)) = 0.14 \pm 0.22$
and $\Delta([Fe/H]) = 0.01 \pm 0.11$. 
Excluding the discrepant M star we obtain:
 $\Delta([Fe/H]) = -0.01 \pm 0.07$.
The deviations reported here are the standard deviations on individual
measurements. 
The temperatures found by \ulyss{} are systematically cooler by 60 K
than those of da Silva, consistently with the
offset mentioned by these authors in their own comparison to the literature.
They found a systematic difference of 39 to 50 K, their measurements being
hotter. The three atmospheric parameters error estimates from our program are
similar to those given by MC simulations and are about 20 times smaller than  
the `external' errors, so we did not put error bars in Fig.~\ref{tgm_omc}, nevertheless 
the deviations (external errors) are identical to those reported by \citet{Silva06}.
We can conclude that the measurements performed with \ulyss{}
are precise and reliable.

\begin{table}
\caption{Stellar atmospheric parameter values for six stars from CFLIB representative of various MK classes.
The atmospheric parameters on the first line are compiled from the literature, and
on the second line fitted by \ulyss.
}
\label{cflib} 
\begin{tabular}{lccccc}
\hline\hline
Name    & Sp.Type & \teff{} & \logg{} & [Fe/H] & Ref.     \\
%\multicolumn{1}{c}{(1)} & (2) & (3) & (4) & (5) & (6)          \\
	&	  & (K)	    &  g.cm$^{-2}$.s$^{-1}$  & (dex)  &		\\
\hline
   HD30614	& O9.5Iae	& 29647	        & 3.05	        &  0.30      & 1  \\
		&		& 33972  	& 3.18		& -0.05	     &  	\\
  HD195324      & A1Ib 	        & 9300	        & 1.90		& -0.11      & 2  \\
    		&		& 9847		& 1.94		& -0.16	     &  \\
  HD114642      & F5.5V 	& 6434	        & 3.83	        & -0.12      & 3  \\
    		&		& 6431		& 4.04		& -0.12	     &  	\\
   HD76151 	& G2V 	        & 5768	        & 4.45	        &  0.06      & 4  \\
     		&		& 5728		& 4.41		&  0.09	     &          \\
   HD10780	& K0V 	        & 5359	        & 4.44	        &  0.02      & 4  \\
     		&		& 5330		& 4.50		&  0.06	     &	        \\
   HD42475 	& M1Iab 	& 4000	        & 0.70		& -0.36      & 5  \\
     		&		& 3988		& 0.32		&  0.02	     &        	\\
\hline
\end{tabular}
Litterature source for the atmospheric parameters.
\begin{itemize}
\item {\sc References} --- (1) \citet{takada77}; (2) \citet{venn95}; (3)
  \citet{takeda07}; (4) \citet{soubiran08}; (5) \citet{luck80}.
\end{itemize}
\end{table}

\begin{figure*}
\centering
\includegraphics{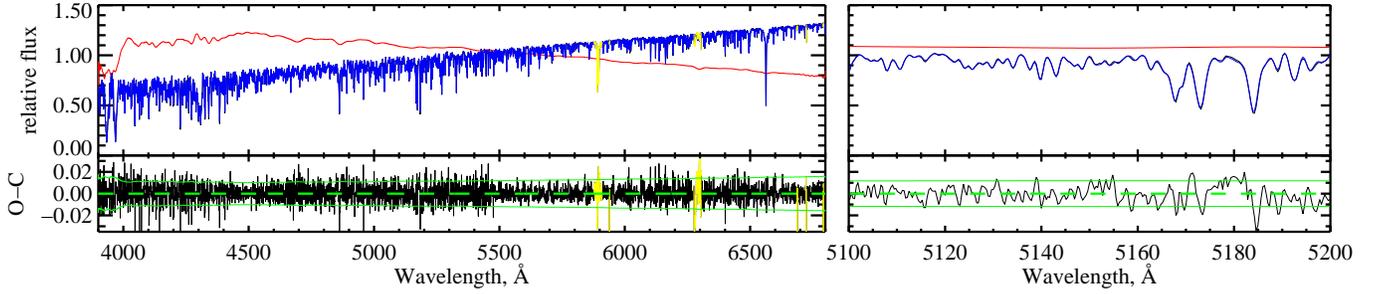}
\caption {Fit of the CFLIB star HD~76151 with a TGM component.
The symbols and conventions are same as in Fig.~\ref{fig:lsf1}.
The order of the multiplicative polynomial is $n=$200.
The right side expands a small wavelength region around Mg$_b$.
}
\label{tgm_fit}
\end{figure*}

In Table~\ref{cflib} we selected six CFLIB stars
representative of the various 
%\citet[][hereafter MK]{william43} 
spectral types and luminosity classes.
%that we will use for the other tests presented later in this paper. 
Fig.~\ref{tgm_fit} presents the fit for a Solar type
star from this list. A detailed discussion of the CFLIB stellar library 
will be made in a separate work.

\subsection{Multiplicative polynomial continuum}
\label{sect:mulpol}

Most stellar analysis programs first require the observed
spectrum to be normalized to a pseudo-continuum, which can be determined either interactively
or automatically. By contrast, \ulyss{} determines this normalization
in the same fitting process by including a multiplicative 
polynomial, $P_n(\lambda)$ in Eq.~\ref{eq:main}, in the model.
This single-step fitting procedure insures that the minimum \csq{}
can be reached and allows to check the possible dependences between this 
continuum and the measured physical parameters.

\begin{figure}
\centering
\includegraphics{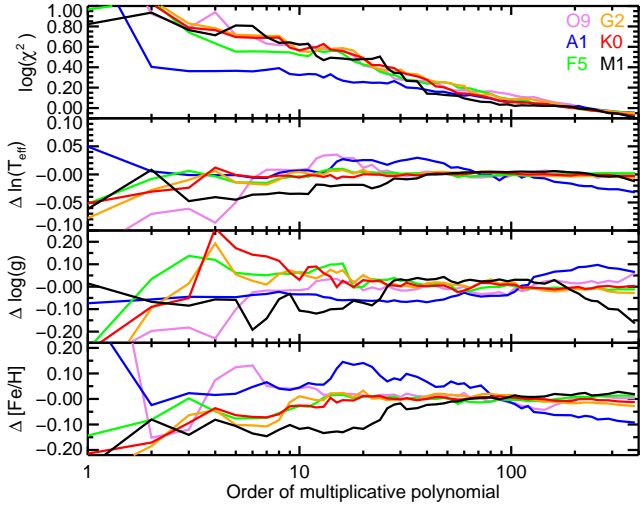}
\caption {The evolution of the stellar atmospheric parameters fit results 
(\csq{}, log(\teff{}), \logg{}, and [Fe/H]) with increasing Legendre 
polynomial degree for 6 example CFLIB stellar spectra. Black, green, red, 
blue, violet and gold colors are respectively standing for each star 
listed in table 1.}
\label{tgm_poly}
\end{figure}

Figure \ref{tgm_poly} presents the results of a series of
fits of the six representative stars from Table~\ref{cflib}
varying the order of $P_n(\lambda)$. 
The observations consist in 8300 independent pixels in
the wavelength range 3900 -- 6800 \AA.
As there is no external estimate of the noise, we
affected a constant weight to all the pixels (except those
rejected as outliers), and computed the noise in order to
reach \csq=1 for $n$=200. We explored the multiplicative polynomial 
order range $0 < n < 800$, while $n$ increasing, the value of the \csq{} 
decreases as a power law.

The atmospheric parameters converge rapidly to their asymptotic values
(defined here as the mean of the solutions for $n > 25$). 
For the F, G,  K and O stars the plateau is reached between
$n = $10  to 15 (the stability of the solution is lower for the O star,
in particular for metallicity). 
For the M star, the plateau is reached at $n$ = 35, but
the fit is not stable above $n$=150.
The A1Ib spectral type CFLIB star HD~195324 displays a significant 
dependence between $n$ and the measured metallicity; it did not
stabilize to a plateau. This is likely due to the 
limited quality of the ELODIE V3.1 interpolator in this under-populated
region of the parameters space. The ELODIE library counts only five 
A-type supergiants and therefore the interpolation is not secure.
Besides, in A-type stars the ELODIE continuum are taken in the flanks
of the Balmer lines and the analysis rely on the multiplicative polynomial
to fit them. We expect that using a flux-calibrated model and an
improved library would improve the situation.

Note, that the variations of the parameters with $n$
are slightly larger than the error bars. On respectively \teff, \logg{}
and [Fe/H] the errors are typically 0.1\%, 0.006 and 0.005
while the dispersion of the values for $n>20$ are 0.2\%,
0.01 and 0.01, i. e. about twice the errors. 
The extremely small internal errors hide some potential biases
of either observational or physical origin. 

Another evidence for the non-degeneracy between the 
atmospheric parameters and $n$, even for values of $n$
considerably larger than what is used in practical cases,
is given by the Monte-Carlo simulations of the next section.
In the case of degeneracy, the error bars computed by the
fitting program would be underestimated.

In order to check if the high values of $n$ are over-fitting the
data (i. e. fit the noise), we made similar experiments with
noise spectra having the same characteristics as the data.
The measured \csq{} decreases as expected, to reach 0.99 for $n = 100$,
0.98 for $n = 400$ and 0.96 for $n = 800$. It is clear that the
\csq{} trend seen for the observation is not due to the over-fitting,
as the slope is much larger. The decrease of \csq{} is probably due 
to the shape of the continuum increasingly better fitted when 
$n$ rises, and to some extent the physical effects ignored by this simple
model. The effect of rotation for hot stars may be the main 
influence, moreover detailed abundances are certainly can not be mimiced by the
multiplicative polynomial.

The importance of this multiplicative polynomial for stellar population studies
is discussed in \citet{koleva08a}. Within the same wavelength range, the fits reach 
a plateau for lower $n$, probably because of the models are flux calibrated.

\subsection{Convergence, \csq{} maps and Monte-Carlo simulations}

\begin{figure}
\includegraphics{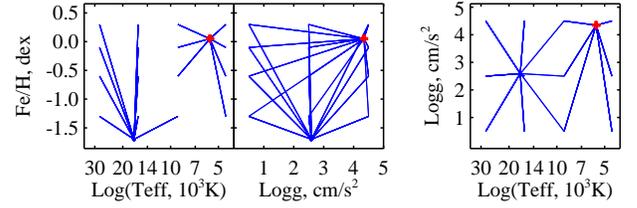}
\caption {Convergence maps on different projections of the 
parameters space for the CFLIB star HD~76151 inverted with the TGM 
component. Red crosses stand for the global minimum solutions found 
by \ulyss{}.} 
\label{tgm_cvg}
\end{figure}

\begin{figure*}
\includegraphics{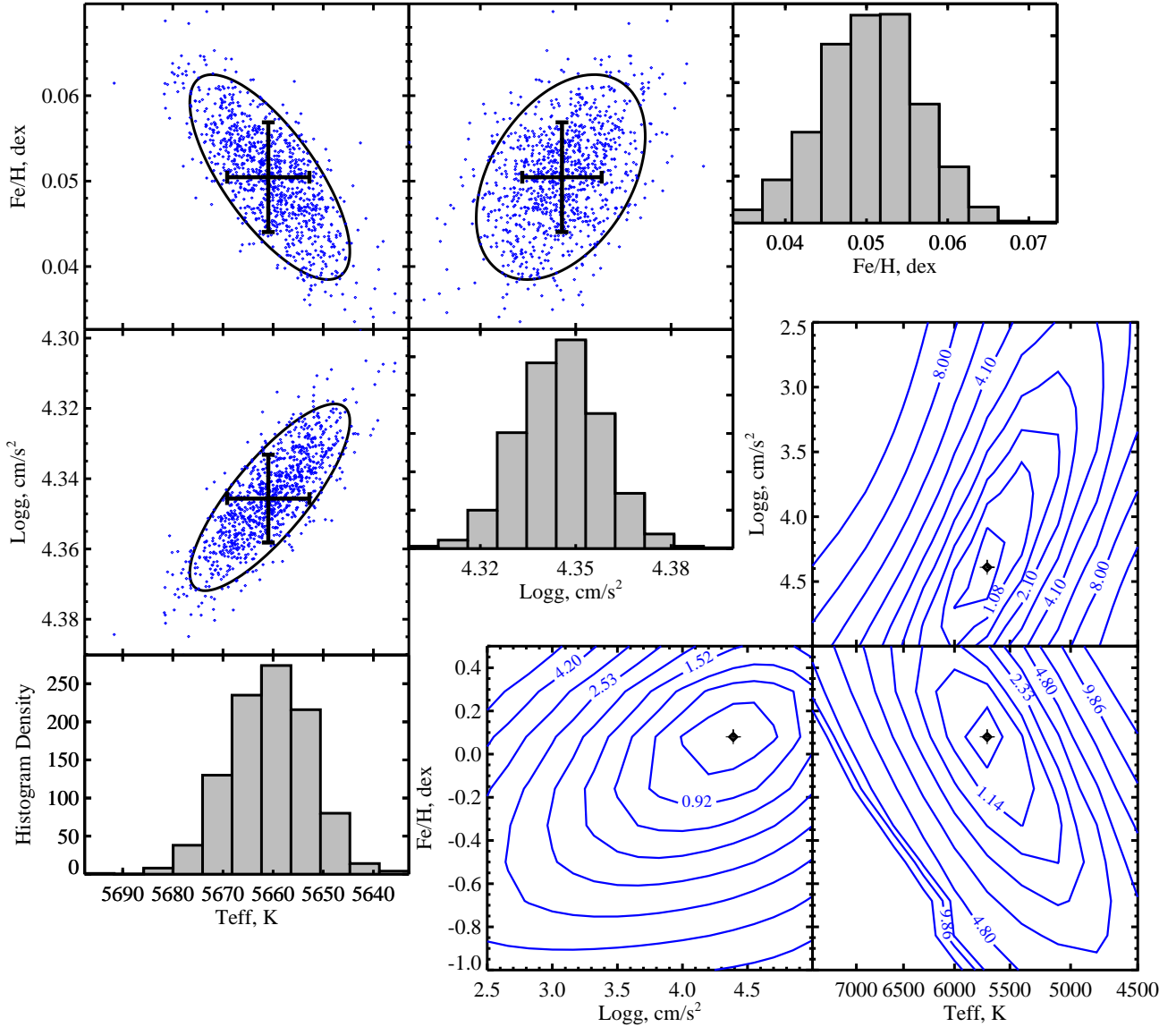}
\caption {Monte-Carlo simulations and \csq{} maps for the CFLIB star HD~76151 
inverted with the ELODIE library as presented in Fig.~\ref{tgm_fit}. 
The three projections of the parameters space are presented.
The 1000 Monte-Carlo simulations are performed adding a random noise 
equivalent to the estimated one.
The overimposed crosses figure the internal errors estimated by \ulyss{}, and the 
ellipses the standard deviation computed from the simulations.} 
\label{tgm_mchistchi}
\end{figure*}

\ulyss{} also includes the tools to assess the significance and validity 
of the results. Fig.~\ref{tgm_cvg} \& \ref{tgm_mchistchi} 
illustrate the usage of convergence maps, Monte-Carlo simulations and 
\csq{} maps to explore the space of parameters.

The convergence map, Fig.~\ref{tgm_cvg}, shows two convergence basins.
In the wide region defined by \teff{} $<$ 10000 K, any choice of 
initial guess will converge toward the correct solution, while hotter
guess temperature may fall into a local minimum in an unphysical 
region. 

Monte-Carlo simulations are used to estimates the errors when the
different parameters are not independent. The errors determined
by Monte-Carlo simulations are compared in Fig.~\ref{tgm_mchistchi}
with those given by the minimization procedure. Though both are in 
approximate agreement, only the simulations can render the effect of the
coupling of the errors between the parameters. The simulations
consist in series of analyses of a spectrum plus a random noise
corresponding to the estimated noise. The added noise has a Gaussian
distribution and take into account the correlation between the pixels
introduced along the processing as stressed by \citet{debruyne03}.
This latter effect is modeled by keeping track of the number of
independent pixels along the steps of the processing, and then
generating a random vector of independent points that is finally resampled
to the actual length of the spectrum.

The \csq{} maps complete the Monte-Carlo simulations by unveiling
the degeneracies and the presence of local minimas. We built them
by choosing a grid of nodes in a 2 dimensional projection
of the parameters space, and performing the minimization for each
node (hence optimizing n-2 parameters). Any local minimum can
be detected, providing that the grid is fine enough. The topology
of these maps also indicates the degeneracies. In the case presented 
by Fig.~\ref{tgm_mchistchi}, measurement of the three atmospheric
parameters for a Solar type star, the maps are regular, showing
weak dependences between the parameters and a single minimum. When
using more complex models, like composite stellar population, the
maps often show local minimas.

\section{History of stellar populations}

By using the SSP component (single stellar population)
provided with \ulyss{}, one can evaluate many 
evolutionary parameters from an integrated spectra. As in the case of the 
TGM component, the SSP component describes the fitting boundaries and the 
overall recipe to create SSP spectra and fit them onto the observed data. 
This time, the CMP is characterised by age, [Fe/H] and [Mg/Fe]
(this last dimension is currently only available in semi-empirical
models under development, see \citealp{prugniel07b}).
A grid of SSPs given in input is spline-interpolated to provide
a continuous function.

The CMPs can be linked to a number of 
population synthesis models. \citet{koleva08a} tested 3 of them:  
Galaxev \citep{bruzual03}, Pegase.HR \citep{leborgne04} and Vazdekis-Miles 
\citep{vazdekis99, sanchezblazquez06}. They concluded that Pegase.HR 
and Vazdekis-Miles are reliable and consistent.

The first step towards reconstructing the star formation history
(SFH) of an object is to calculate 
its SSP-\emph{equivalent} parameters by using a single CMP
that interpolate a grid of SSP in age, [Fe/H] and possibly [Mg/Fe].
These SSP-equivalent parameters correspond (more or less) 
to the luminosity weighted average over all the population that can
have a distribution in age and metallicity. 
The present analysis program has been used by \citet{koleva08a} who 
has shown that reliable information can be retrieved.
The metallicity of Galactic globular clusters can be compared
to the determinations derived from spectroscopy of individual stars
with a precision of 0.1 dex, which is the actual precision of these
latter measurements.

If the object has a complex SFH, with several epochs of star formation,
SSP-\emph{equivalent} parameters are essentially representative of the
star formation burst that dominates the light (often the more recent).
The \ulyss{} package can be used to reconstruct a detailed SFH, in practice
limited to 2 to 4 epochs of star formation (see also VESPA; \citealp{VESTA}).

\subsection{Complex Stellar Population: Application to NGC205}
The galaxies have in general a complex SFH and retracing the star formation rate
along the history is a fundamental information for the physics of the galaxies
and for the cosmology.
In principle, one can access such information by fitting 
directly a positive linear combination of many SSPs, but such approach
would be unstable because of the degeneracies between the components
and require a regularization.

To circumvent these degeneracies, we start with simple physical 
assumptions, like the presence of an old stellar population, then 
divide the time axis in intervals (by setting limits in two or more
intervals). As the number of free parameters raise, local minima
appear and global minimization are required; the \csq{} maps
help to understand the structure of the parameters space.
Usually the fit is performed several times, with increasing 
number of components and varying limits on the population boxes. Then, 
doing Monte-Carlo simulations
and checking the residuals of the fits, it is possible 
to assess the relevance of the solutions and select the most probable SFH. 

As an example, we analyse the star formation in the inner 2\arcsec{}
(roughly the size of the nucleus, \citealt{butler05}) of 
NGC~205. The data, taken from  \citet{simien02}, have S/N $\approx$ 50
in the central region and spectral resolution of R\,$\sim$\,5000. 
For this present demonstration, we analyse this spectrum in terms of
two epochs of star formation (\ie{} 
2 CMPs, each one being an SSP): one `young' (age $<$800\,Myr) and one 
`old' (800\,$<$\,age\,$<$\,14\,000\,Myr).
This hypothesis is not bound to an a priori knowledge of the stellar population;
it is essentially a choice of time resolution. Depending on S/N, resolution
and wavelength range, the number of components may be increased;
\eg{} in \citet{koleva09} the same data are decomposed in four epochs of star formation.

%\begin{table*}[tbh]
%\caption{Ages and metallicities of the central $2''$ of NGC205.
%from one and two bursts fit. The error estimations on the fractions
%are computed from MC simulations.}
%\begin{tabular}{lccccccc}
%\hline
%\hline
%n fitted SSPs & \csq{} & $age_0$, Gyr & $[Fe/H]_0$, dex & Fraction$_0$ (Light fraction) & 
%$age_1$, Gyr & $[Fe/H]_1, dex$ & Fraction$_1$ (Light fraction) \\
%\hline
%1 &  0.94 & & &    &  1.26 $\pm$ 0.01 &  -0.65 $\pm$ 0.02 & 0.99 \\
%2 & 0.82 & 0.13 $\pm$ 0.02 & 0.28 $\pm$ 0.09 & 0.08 $\pm$ 0.02 (25 $\pm$ 7.7~\%) &
%1.81 $\pm$ 0.01 & -0.73 $\pm$ 0.01 & 1.13 $\pm$ 0.08 (75 $\pm$ 5~\%)\\
%\hline
%\end{tabular}
%\label{tab:res}
%\end{table*}
%
%
%
\begin{table}[tb]
\caption{Ages and metallicities of the central $2''$ of NGC205.
The first part of the table give the direct fits, for one and two SSPs.
The second part give the mean values and standard deviations derived
from 1000 Monte Carlo simulations.
In this second part, MC1 is the result of the Monte-Carlo analysis
of the SSP direct solution with a 2-SSP models.
}
\begin{tabular}{lccccc}
\hline
\hline
SSP 	& \csq{}	& $age$ 	& [Fe/H]	&  $f_{mass}$	& $f_{light}$	\\
	& 		    & (Gyr)		& (dex)		& 		&		\\
\hline
1 / 1	& 1.37		        & 1.27$\pm$0.02 & -0.67$\pm$0.02 & - 		& -  		\\
\hline
1 / 1	& MC                    & 1.27$\pm$0.02 & -0.67$\pm$0.02 & -             & -            \\
\hline
1 / 2	& \multirow{2}{*}{1.22} & 0.13$\pm$0.02 & 0.29$\pm$0.07  & 0.06$\pm$0.00 & 0.25$\pm$0.00 \\
2 / 2  &  		        & 1.81$\pm$0.08 & -0.73$\pm$0.01 & 0.94$\pm$0.00 & 0.75$\pm$0.00 \\
\hline
1 / 2	& \multirow{2}{*}{MC}   & 0.14$\pm$0.04 &  0.28$\pm$0.04 & 0.07$\pm$0.02 & 0.26$\pm$0.04\\
2 / 2  &  		        & 1.90$\pm$0.36 & -0.74$\pm$0.08 & 0.93$\pm$0.07 & 0.74$\pm$0.04\\
\hline
1 / 2	&  \multirow{2}{*}{MC1} & -             & -              & -             & -            \\
2 / 2	&                       & 1.30$\pm$0.03 & -0.67$\pm$0.04 & -             & -            \\
\hline
\end{tabular}
\label{tab:res}
\end{table}

\begin{figure}[tbh]
\begin{center}
\includegraphics[width=0.5\textwidth]{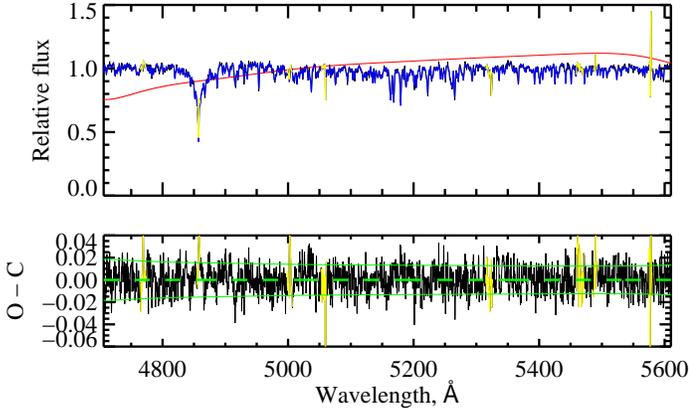} 
\caption{Fit result for the central 2\second{} of NGC~205 with 2 SSP components. In the \emph{top} panel the \emph{black} line represents the observed spectra, the \emph{blue} line is the fitted model and the \emph{red} line shows the multiplicative polynomial ($P_n(\lambda)$). The \emph{bottom} panel shows the residuals (\emph{black}), mean (\emph{green dashed}) and 1$\sigma$ deviation (\emph{solid green}). In both panels, some of the data were not considered for the fit (\emph{yellow}).}
\label{fig:fit}
\end{center}
\end{figure}

\begin{figure}[tbh]
\begin{center}
\includegraphics[width=8.5cm]{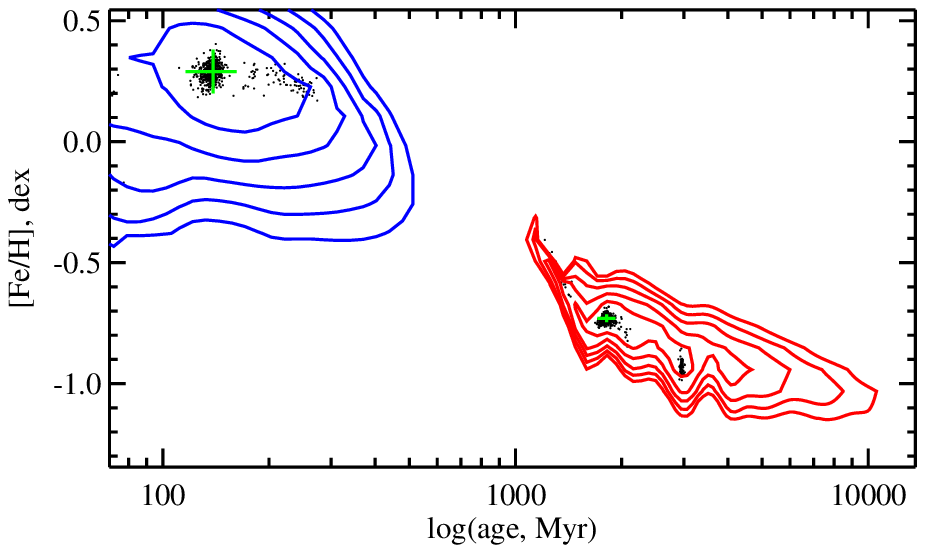}
\caption{Monte-Carlo simulation (\emph{dots}) and \csq{} map (\emph{contours}) for a 2 CMP fit to spectra of the inner 2\arcsec{} of NGC~205. The \emph{blue contours} represents the \csq{} levels for the young stellar population (age constrained to be $<800$ Myr) and the \emph{red contours} are for the old population (age between 800 and 14\,000 Myr). 
}
\label{fig:mc}
\end{center}
\end{figure}

Figure \ref{fig:fit} and Table \ref{tab:res} shows the results of our 
analysis. For the young component we find an age of $\sim$130~Myr 
consistent with photometric results (J, H, K photometry) from 
\citet{davidge03} and with \citet{cepa88} who found from orbital considerations 
that NGC~205 crossed the disk of M31 about 100~Myr ago.
It represents about 25\% of the light and only about 7\% of the mass. 

The Monte-Carlo simulations of Fig.~\ref{fig:mc} show that
we can distinguish between the two populations, in the sense 
that the two clouds corresponding to these populations
are well separated.
However, the existence of two bursts was one of our hypotheses
and to test its validity we will apply the same analysis to the best fit SSP 
of the first NGC~205 experiment.
We find that a young burst would be detected in about 10\% 
of the cases in Monte-Carlo simulations, but it is easily rejected as the 
solution does not cluster around a marked minimum.

The mean solutions estimated from the Monte-Carlo simulations 
with two bursts, given in Table~\ref{tab:res} (lines noted 'MC' in the \csq{} column), 
are compatible with the direct
fit, but the errors are significantly larger (because of the degeneracies).
It is also interesting to see that
the analysis with a 2-bursts model of the best-fit SSP produce an unbiased solution
(lines noted 'MC1').

Examining Fig.~\ref{fig:mc} in more details, we note that the distribution of 
the Monte-Carlo solutions
are not simple Gaussians centered on the direct solution.
For the old burst, there is a small and concentrated secondary cloud 
gathering 10\% of the solutions at an age of about 3 Gyr and [Fe/H]=-1.
The solutions belonging to that cloud form also a tail at the old
side of the young burst.
This feature is associated to a local minimum detected on the \csq{}
map whose depth is almost similar to the solution. Because of the random
noise, this local minimum can become the actual solution.
It means that the two minima cannot be statistically distinguished.
The morphology of the map gives the impression that
the oldest cloud is an artifact due to a discontinuity
of the model (either in the population model of interpolation
of the Elodie spectra). But there is no objective argument
to reject this (less probable) solution.

\subsection{Exponentially declining or constant SFR populations}

Our last series of experiments consists in analysing simulated populations
with an extended star formation history. We consider here two scenarios:
a constant star formation rate (SFR) stopped 500\,Myr ago and an 
exponentially declining 
SFR with a characteristic time, $\tau$, of 5 Gyr. These spectra 
were computed with Pegase.HR and ELODIE.3.1.
In both cases we analysed the simulated galaxies with a model made of 3 CMPs
defined by the age limits: [12,800], [800, 5000], [12000] Myr.
The goal is to see if we can distinguish between different scenarios
of star formation.

\begin{figure}
\includegraphics{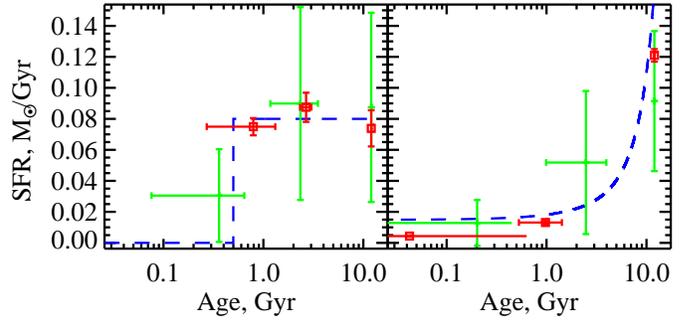}
\caption {Star formation histories of two simulated populations
with constant (upper panel) and exponentially decreasing
(lower panel) star formation rates. The vertical axis is the SFR normalized to
a total mass of formed stars equal to 1 M$_\odot$. 
The blue dashed lines represents the 
simulated star formation. The red squares are the direct solution of the fit to
a 3 SSPs model. The green crosses are the results from MC
simulations (200 inversions with S/N=50). 
}
\label{fig-diff_SFH}
\end{figure}

We fitted the direct solutions (without noise) and we
made a series of 200 inversions with different realization 
of the added noise corresponding to S/N=50. 
 Fig.\ref{fig-diff_SFH} presents the results 
as the SFR versus time.
For both simulated populations we find significant contribution
(more than 10\% in light) in the youngest component box, where
the star formation had the lowest rate. 
The SFR reconstructions reproduce well the simulated histories
and the MC simulations agree with the direct solution within the 
uncertainties. The direct solution underestimates the error bars
due to the degeneracies between the ages and the relative weights of
each component.

The error bars on the age are larger than in the case of NGC~205.
In fact the individual solutions of the MC realizations span
the whole age range allowed for each of the components. This may be
an indication that by contrast, the star formation history of NGC~205
is marked by a recent short burst rather than a smoothly declining SFR.

\section{Conclusion}

We presented the \ulyss{} packages and its applications to the
analysis of stellar atmosphere and stellar population spectra.
This packages is simple to use and is an efficient tool
to determine the atmospheric parameters of stars (\teff, \logg{}
and [Fe/H]). The convergence region and the degeneracies 
can be studied in details and the errors on estimated
parameters are robustly determined.
\ulyss{} is also used to recover the history of star formation in galaxies and stellar clusters by decomposing the observed spectrum as a series of SSPs.
The simultaneous analysis of the kinematics
and of the stellar mix of a population allows to break degeneracies
and gain in reliability and precision on both the kinematics
and the star formation history.

\ulyss{} is available for download (\url{http://ulyss.univ-lyon1.fr}).
Beside the applications described in the present paper, it contains
other {\it components} (e. g. LINE, used to fit emission lines)
and can easily be extended to other applications.

\begin{acknowledgements}
MK acknowledges a grant from the French embassy in Sofia and
YW acknowledges a grant from China Scholarship Council.
We are grateful to Craig Markwardt (MPFIT) and Michele Cappellari 
(pPXF \& BVLS) who freely distribute IDL/GDL routines which made this 
project possible. We thank also David Fanning (graphic library) and 
the contributors to the IDL Astronomical library. We acknowledge the 
students and technicians who helped to the development and tests of the method and of 
the present code over the years: Maela Collobert, Nicolas Bavouzet, Igor 
Chilingarian, Paul Blond\'e and Martin France.
We thank the anonymous referee for constructive comments that helped to
improve the manuscript.

\end{acknowledgements}

\bibliographystyle{aa} % style aa.bst 
\bibliography{ulyss}   % bibtex database 

\begin{thebibliography}{65}
\expandafter\ifx\csname natexlab\endcsname\relax\def\natexlab#1{#1}\fi

\bibitem[{{Bailer-Jones} {et~al.}(1997){Bailer-Jones}, {Irwin}, {Gilmore}, \&
  {von Hippel}}]{bailer-jones97}
{Bailer-Jones}, C.~A.~L., {Irwin}, M., {Gilmore}, G., \& {von Hippel}, T. 1997,
  \mnras, 292, 157

\bibitem[{{Bender}(1990)}]{bender90}
{Bender}, R. 1990, \aap, 229, 441

\bibitem[{{Bolstad}(1977)}]{bolstad77}
{Bolstad}, J. 1977, VARPRO-A general non-linear least-squares fitting code
  (Stanford Univ.)

\bibitem[{Bouchard {et~al.}(2009)Bouchard, Koleva, \& Prugniel}]{bouchard09}
Bouchard, A., Koleva, M., \& Prugniel, P. 2009, in preparation

\bibitem[{{Bruzual} \& {Charlot}(2003)}]{bruzual03}
{Bruzual}, G. \& {Charlot}, S. 2003, \mnras, 344, 1000

\bibitem[{{Butler} \& {Mart{\'{\i}}nez-Delgado}(2005)}]{butler05}
{Butler}, D.~J. \& {Mart{\'{\i}}nez-Delgado}, D. 2005, \aj, 129, 2217

\bibitem[{{Cappellari} \& {Emsellem}(2004)}]{cappellari04}
{Cappellari}, M. \& {Emsellem}, E. 2004, \pasp, 116, 138

\bibitem[{{Cayrel de Strobel} {et~al.}(2001){Cayrel de Strobel}, {Soubiran}, \&
  {Ralite}}]{cayrel01}
{Cayrel de Strobel}, G., {Soubiran}, C., \& {Ralite}, N. 2001, \aap, 373, 159

\bibitem[{{Cepa} \& {Beckman}(1988)}]{cepa88}
{Cepa}, J. \& {Beckman}, J.~E. 1988, \aap, 200, 21

\bibitem[{{Cid Fernandes} {et~al.}(2005){Cid Fernandes}, {Mateus}, {Sodr{\'e}},
  {Stasi{\'n}ska}, \& {Gomes}}]{cidfernandes05}
{Cid Fernandes}, R., {Mateus}, A., {Sodr{\'e}}, L., {Stasi{\'n}ska}, G., \&
  {Gomes}, J.~M. 2005, \mnras, 358, 363

\bibitem[{{Coelho} {et~al.}(2005){Coelho}, {Barbuy}, {Mel{\'e}ndez},
  {Schiavon}, \& {Castilho}}]{coelho05}
{Coelho}, P., {Barbuy}, B., {Mel{\'e}ndez}, J., {Schiavon}, R.~P., \&
  {Castilho}, B.~V. 2005, \aap, 443, 735

\bibitem[{{da Silva} {et~al.}(2006){da Silva}, {Girardi}, {Pasquini},
  {Setiawan}, {von der L{\"u}he}, {de Medeiros}, {Hatzes}, {D{\"o}llinger}, \&
  {Weiss}}]{Silva06}
{da Silva}, L., {Girardi}, L., {Pasquini}, L., {et~al.} 2006, \aap, 458, 609

\bibitem[{{Davidge}(2003)}]{davidge03}
{Davidge}, T.~J. 2003, \apj, 597, 289

\bibitem[{{de Bruyne} {et~al.}(2003){de Bruyne}, {Vauterin}, {de Rijcke}, \&
  {Dejonghe}}]{debruyne03}
{de Bruyne}, V., {Vauterin}, P., {de Rijcke}, S., \& {Dejonghe}, H. 2003,
  \mnras, 339, 215

\bibitem[{{Franx} {et~al.}(1989){Franx}, {Illingworth}, \& {Heckman}}]{franx89}
{Franx}, M., {Illingworth}, G., \& {Heckman}, T. 1989, \apj, 344, 613

\bibitem[{Golub \& Pereyra(1973)}]{golub73}
Golub, G. \& Pereyra, V. 1973, SIAM, 10, 413

\bibitem[{{Heavens} {et~al.}(2000){Heavens}, {Jimenez}, \& {Lahav}}]{MOPED}
{Heavens}, A.~F., {Jimenez}, R., \& {Lahav}, O. 2000, \mnras, 317, 965

\bibitem[{{Hekker} \& {Mel{\'e}ndez}(2007)}]{Hekker07}
{Hekker}, S. \& {Mel{\'e}ndez}, J. 2007, \aap, 475, 1003

\bibitem[{{Katz} {et~al.}(1998){Katz}, {Soubiran}, {Cayrel}, {Adda}, \&
  {Cautain}}]{katz98}
{Katz}, D., {Soubiran}, C., {Cayrel}, R., {Adda}, M., \& {Cautain}, R. 1998,
  \aap, 338, 151

\bibitem[{{Koleva}(2009)}]{koleva09}
{Koleva}, M. 2009, PhD thesis, University of Lyon

\bibitem[{{Koleva} {et~al.}(2007){Koleva}, {Bavouzet}, {Chilingarian}, \&
  {Prugniel}}]{koleva07}
{Koleva}, M., {Bavouzet}, N., {Chilingarian}, I., \& {Prugniel}, P. 2007, in
  Science Perspectives for 3D Spectroscopy, ed. M.~{Kissler-Patig}, J.~R.
  {Walsh}, \& M.~M. {Roth}, 153--+

\bibitem[{{Koleva} {et~al.}(2008{\natexlab{a}}){Koleva}, {Gupta}, {Prugniel},
  \& {Singh}}]{koleva08b}
{Koleva}, M., {Gupta}, R., {Prugniel}, P., \& {Singh}, H. 2008{\natexlab{a}},
  in Astronomical Society of the Pacific Conference Series, Vol. 390, Pathways
  Through an Eclectic Universe, ed. J.~H. {Knapen}, T.~J. {Mahoney}, \&
  A.~{Vazdekis}, 302--+

\bibitem[{{Koleva} {et~al.}(2008{\natexlab{b}}){Koleva}, {Prugniel}, \& {De
  Rijcke}}]{koleva08c}
{Koleva}, M., {Prugniel}, P., \& {De Rijcke}, S. 2008{\natexlab{b}},
  Astronomische Nachrichten, 329, 968

\bibitem[{{Koleva} {et~al.}(2008{\natexlab{c}}){Koleva}, {Prugniel}, {Ocvirk},
  {Le Borgne}, \& {Soubiran}}]{koleva08a}
{Koleva}, M., {Prugniel}, P., {Ocvirk}, P., {Le Borgne}, D., \& {Soubiran}, C.
  2008{\natexlab{c}}, \mnras, 385, 1998

\bibitem[{Lawson \& Hanson(1995)}]{lawson95}
Lawson, C.~L. \& Hanson, R.~J. 1995, Solving Least Squares Problems, Classics
  in Applied Mathematics No.~15 (Philadelphia, Penn.: SIAM)

\bibitem[{{Le Borgne} {et~al.}(2004){Le Borgne}, {Rocca-Volmerange},
  {Prugniel}, {Lan{\c c}on}, {Fioc}, \& {Soubiran}}]{leborgne04}
{Le Borgne}, D., {Rocca-Volmerange}, B., {Prugniel}, P., {et~al.} 2004, \aap,
  425, 881

\bibitem[{{Luck} \& {Bond}(1980)}]{luck80}
{Luck}, R.~E. \& {Bond}, H.~E. 1980, \apj, 241, 218

\bibitem[{{Marquart}(1963)}]{marquardt63}
{Marquart}, D.~W. 1963, SIAM, 11, 431

\bibitem[{{Michielsen} {et~al.}(2007){Michielsen}, {Koleva}, {Prugniel},
  {Zeilinger}, {De Rijcke}, {Dejonghe}, {Pasquali}, {Ferreras}, \&
  {Debattista}}]{michielsen07}
{Michielsen}, D., {Koleva}, M., {Prugniel}, P., {et~al.} 2007, \apjl, 670, L101

\bibitem[{Mor\'e {et~al.}(1980)Mor\'e, Garbow, \& Hillstrom}]{more80}
Mor\'e, J.~J., Garbow, B.~S., \& Hillstrom, K.~E. 1980, User Guide for
  {MINPACK-1}, Report ANL-80-74, ANL, ANL

\bibitem[{{Moultaka}(2005)}]{moultaka05}
{Moultaka}, J. 2005, \aap, 430, 95

\bibitem[{{Munari} {et~al.}(2005){Munari}, {Sordo}, {Castelli}, \&
  {Zwitter}}]{munari05}
{Munari}, U., {Sordo}, R., {Castelli}, F., \& {Zwitter}, T. 2005, \aap, 442,
  1127

\bibitem[{{Neilson} \& {Lester}(2008)}]{Neilson08}
{Neilson}, H.~R. \& {Lester}, J.~B. 2008, \aap, 490, 807

\bibitem[{{Ocvirk} {et~al.}(2006{\natexlab{a}}){Ocvirk}, {Pichon}, {Lan{\c
  c}on}, \& {Thi{\'e}baut}}]{ocvirk06b}
{Ocvirk}, P., {Pichon}, C., {Lan{\c c}on}, A., \& {Thi{\'e}baut}, E.
  2006{\natexlab{a}}, \mnras, 365, 74

\bibitem[{{Ocvirk} {et~al.}(2006{\natexlab{b}}){Ocvirk}, {Pichon}, {Lan{\c
  c}on}, \& {Thi{\'e}baut}}]{ocvirk06a}
{Ocvirk}, P., {Pichon}, C., {Lan{\c c}on}, A., \& {Thi{\'e}baut}, E.
  2006{\natexlab{b}}, \mnras, 365, 46

\bibitem[{{Panter} {et~al.}(2003){Panter}, {Heavens}, \& {Jimenez}}]{panter03}
{Panter}, B., {Heavens}, A.~F., \& {Jimenez}, R. 2003, \mnras, 343, 1145

\bibitem[{{Prugniel} {et~al.}(2005){Prugniel}, {Chilingarian}, \&
  {Popovi{\'c}}}]{Chilingarian05}
{Prugniel}, P., {Chilingarian}, I., \& {Popovi{\'c}}, L.~{\v C}. 2005, Memorie
  della Societa Astronomica Italiana Supplement, 7, 42

\bibitem[{{Prugniel} \& {Koleva}(2007)}]{Prugniel07c}
{Prugniel}, P. \& {Koleva}, M. 2007, in American Institute of Physics
  Conference Series, Vol. 938, Spectral Line Shapes in Astrophysics, ed. L.~C.
  {Popovic} \& M.~S. {Dimitrijevic}, 27--34

\bibitem[{{Prugniel} {et~al.}(2007{\natexlab{a}}){Prugniel}, {Koleva},
  {Ocvirk}, {Le Borgne}, \& {Soubiran}}]{prugniel07b}
{Prugniel}, P., {Koleva}, M., {Ocvirk}, P., {Le Borgne}, D., \& {Soubiran}, C.
  2007{\natexlab{a}}, in IAU Symposium, Vol. 241, IAU Symposium, ed.
  A.~{Vazdekis} \& R.~F. {Peletier}, 68--72

\bibitem[{Prugniel {et~al.}(2009)Prugniel, Singh, Gupta, Wu, \&
  Koleva}]{prugniel09}
Prugniel, P., Singh, H.~P., Gupta, R., Wu, Y., \& Koleva, M. 2009, in
  preparation

\bibitem[{{Prugniel} \& {Soubiran}(2001)}]{prugniel01}
{Prugniel}, P. \& {Soubiran}, C. 2001, \aap, 369, 1048

\bibitem[{{Prugniel} {et~al.}(2007{\natexlab{b}}){Prugniel}, {Soubiran},
  {Koleva}, \& {Le Borgne}}]{prugniel07}
{Prugniel}, P., {Soubiran}, C., {Koleva}, M., \& {Le Borgne}, D.
  2007{\natexlab{b}}, ArXiv Astrophysics e-prints, arXiv:astro-ph/0703658

\bibitem[{{Recio-Blanco} {et~al.}(2006){Recio-Blanco}, {Bijaoui}, \& {de
  Laverny}}]{MATISSE}
{Recio-Blanco}, A., {Bijaoui}, A., \& {de Laverny}, P. 2006, \mnras, 370, 141

\bibitem[{{Rix} \& {White}(1992)}]{rix92}
{Rix}, H.-W. \& {White}, S.~D.~M. 1992, \mnras, 254, 389

\bibitem[{{S{\'a}nchez-Bl{\'a}zquez} {et~al.}(2006){S{\'a}nchez-Bl{\'a}zquez},
  {Peletier}, {Jim{\'e}nez-Vicente}, {Cardiel}, {Cenarro},
  {Falc{\'o}n-Barroso}, {Gorgas}, {Selam}, \& {Vazdekis}}]{sanchezblazquez06}
{S{\'a}nchez-Bl{\'a}zquez}, P., {Peletier}, R.~F., {Jim{\'e}nez-Vicente}, J.,
  {et~al.} 2006, \mnras, 371, 703

\bibitem[{{Sargent} {et~al.}(1977){Sargent}, {Schechter}, {Boksenberg}, \&
  {Shortridge}}]{sargent77}
{Sargent}, W.~L.~W., {Schechter}, P.~L., {Boksenberg}, A., \& {Shortridge}, K.
  1977, \apj, 212, 326

\bibitem[{{Sarzi} {et~al.}(2006){Sarzi}, {Falc{\'o}n-Barroso}, {Davies},
  {Bacon}, {Bureau}, {Cappellari}, {de Zeeuw}, {Emsellem}, {Fathi},
  {Krajnovi{\'c}}, {Kuntschner}, {McDermid}, \& {Peletier}}]{sarzi06}
{Sarzi}, M., {Falc{\'o}n-Barroso}, J., {Davies}, R.~L., {et~al.} 2006, \mnras,
  366, 1151

\bibitem[{{Shkedy} {et~al.}(2007){Shkedy}, {Decin}, {Molenberghs}, \&
  {Aerts}}]{shkedy07}
{Shkedy}, Z., {Decin}, L., {Molenberghs}, G., \& {Aerts}, C. 2007, \mnras, 377,
  120

\bibitem[{{Simien} \& {Prugniel}(2002)}]{simien02}
{Simien}, F. \& {Prugniel}, P. 2002, \aap, 384, 371

\bibitem[{{Snider} {et~al.}(2001){Snider}, {Allende Prieto}, {von Hippel},
  {Beers}, {Sneden}, {Qu}, \& {Rossi}}]{snider01}
{Snider}, S., {Allende Prieto}, C., {von Hippel}, T., {et~al.} 2001, \apj, 562,
  528

\bibitem[{{Soubiran} {et~al.}(2008){Soubiran}, {Bienaym{\'e}}, {Mishenina}, \&
  {Kovtyukh}}]{soubiran08}
{Soubiran}, C., {Bienaym{\'e}}, O., {Mishenina}, T.~V., \& {Kovtyukh}, V.~V.
  2008, \aap, 480, 91

\bibitem[{{Soubiran} {et~al.}(1998){Soubiran}, {Katz}, \&
  {Cayrel}}]{soubiran98}
{Soubiran}, C., {Katz}, D., \& {Cayrel}, R. 1998, \aaps, 133, 221

\bibitem[{{Sousa} {et~al.}(2007){Sousa}, {Santos}, {Israelian}, {Mayor}, \&
  {Monteiro}}]{sousa07}
{Sousa}, S.~G., {Santos}, N.~C., {Israelian}, G., {Mayor}, M., \& {Monteiro},
  M.~J.~P.~F.~G. 2007, \aap, 469, 783

\bibitem[{{Takada}(1977)}]{takada77}
{Takada}, M. 1977, \pasj, 29, 439

\bibitem[{{Takeda}(2007)}]{takeda07}
{Takeda}, Y. 2007, \pasj, 59, 335

\bibitem[{{Tojeiro} {et~al.}(2007){Tojeiro}, {Heavens}, {Jimenez}, \&
  {Panter}}]{VESTA}
{Tojeiro}, R., {Heavens}, A.~F., {Jimenez}, R., \& {Panter}, B. 2007, \mnras,
  381, 1252

\bibitem[{{Tonry} \& {Davis}(1979)}]{tonry79}
{Tonry}, J. \& {Davis}, M. 1979, \aj, 84, 1511

\bibitem[{{Valdes} {et~al.}(2004){Valdes}, {Gupta}, {Rose}, {Singh}, \&
  {Bell}}]{valdes04}
{Valdes}, F., {Gupta}, R., {Rose}, J.~A., {Singh}, H.~P., \& {Bell}, D.~J.
  2004, \apjs, 152, 251

\bibitem[{{van der Marel} \& {Franx}(1993)}]{marel93}
{van der Marel}, R.~P. \& {Franx}, M. 1993, \apj, 407, 525

\bibitem[{{Vazdekis}(1999)}]{vazdekis99}
{Vazdekis}, A. 1999, \apj, 513, 224

\bibitem[{{Vazdekis} {et~al.}(2003){Vazdekis}, {Cenarro}, {Gorgas}, {Cardiel},
  \& {Peletier}}]{vazdekis03}
{Vazdekis}, A., {Cenarro}, A.~J., {Gorgas}, J., {Cardiel}, N., \& {Peletier},
  R.~F. 2003, \mnras, 340, 1317

\bibitem[{{Venn}(1995)}]{venn95}
{Venn}, K.~A. 1995, \apjs, 99, 659

\bibitem[{{Willemsen} {et~al.}(2005){Willemsen}, {Hilker}, {Kayser}, \&
  {Bailer-Jones}}]{willemsen05}
{Willemsen}, P.~G., {Hilker}, M., {Kayser}, A., \& {Bailer-Jones}, C.~A.~L.
  2005, \aap, 436, 379

\bibitem[{{Wittkowski} {et~al.}(2006){Wittkowski}, {Hummel}, {Aufdenberg}, \&
  {Roccatagliata}}]{Wittkowski06}
{Wittkowski}, M., {Hummel}, C.~A., {Aufdenberg}, J.~P., \& {Roccatagliata}, V.
  2006, \aap, 460, 843

\bibitem[{Wu(2009)}]{wu09}
Wu, Y. 2009, in preparation

\end{thebibliography}
\end{document}